\documentclass[useAMS,usenatbib]{mn2e}
\usepackage{graphicx}
\usepackage[T1]{fontenc}
\usepackage{ae,aecompl}

\usepackage{amsmath}	% Advanced maths commands
\usepackage{amssymb}	% Extra maths symbols

\usepackage{color}

\newcommand{\msun}{${\cal M}_\odot$\,}

\title[Eccentricity distribution]{Eccentricity distribution of wide low-mass binaries}

\author[Tokovinin]{
Andrei~Tokovinin\thanks{E-mail: atokovinin@ctio.noao.edu} \\
Cerro Tololo Inter-American Observatory/ NSF's National
Optical-Infrared Astronomical Research Laboratory \\
Casilla 603, La Serena, Chile\\
}

\begin{document}

\date{-}

\pagerange{\pageref{firstpage}--\pageref{lastpage}} \pubyear{2020}

\maketitle

\label{firstpage}

\begin{abstract}
Distribution of  eccentricities of very  wide (up to 10  kau) low-mass
binaries in  the solar  neighborhood is studied  using the  catalogue of
El-Badry and  Rix (2018) based on  {\it Gaia}. Direction  and speed of
relative motions in wide  pairs contain statistical information on the
eccentricity distribution,  otherwise inaccessible owing  to very long
orbital periods.   It is found  that the eccentricity  distribution is
close to  the linear  (thermal) one $f(e)  = 2e$. However,  pairs with
projected  separations $<$200  au  have less  eccentric orbits,  while
$f(e)$ for  wide pairs with $s > 1 $ kau  appears to be slightly super-thermal, with
an excess of  very eccentric orbits.  Eccentricity of  any wide binary
can  be constrained  statistically using  direction and  speed  of its
motion. The thermal eccentricity distribution signals an important role of
the stellar  dynamics   in  the  formation  of   wide  binaries,  although
disk-assisted  capture  also can  produce  such  pairs with  eccentric
orbits.
\end{abstract}


%%----

%----------------------------------------------------------
\section{Introduction}
\label{sec:intro}

Formation of  binaries is an  actively debated subject. It  is already
clear that multiple systems are formed by several different mechanisms
and  that  binary  statistics  depend on  the  environment.   However,
quantitative and predictive models are still lacking.

Orbits of wide binaries are a fossil record of their formation process
and  early  dynamical  evolution.   In  this  note,  I  focus  on  the
eccentricity distribution. Dynamical interactions in a cluster are
expected to  produce
a ``thermal'' eccentricity distribution $f(e)=2e$,  although this
state might not be reached in actual clusters
\citep{Geller2019}.  Decay of chaotic triple stars leads to a slightly super-thermal
eccentricity distribution \citep{Stone2019},  while eccentricities of binaries
surviving interactions with other stars (scattering) are distributed thermally
\citep{Antognini2016}. 
 On the other hand,  dissipative forces
(tides  or gas  friction) tend to decrease  the eccentricity.   Ejections from
unstable  triples, on the  contrary, produce  wide binaries  with very
eccentric orbits and a ``super-thermal'' $f(e)$ \citep{Reipurth2012}.

Here  a   catalogue  of  wide   binaries  within  200\,pc   compiled  by
\citet[][hereafter ER2018]{EB2018}  on the basis of  {\it Gaia} second
data  release  \citep{Gaia}   is  used.   Accurate  astrometry  allows
measurement of relative motion in the nearby resolved pairs which, in
turn, contains  some information on  the eccentricity.  \citet{TK2016}
explored  this  option using  classical  (pre-{\it Gaia})  data on  binaries
within 67\,pc with  separations from 50 to a few  hundred au.  The new
{\it  Gaia} sample  of  wide  binaries is  larger  and more  accurate,
allowing us to reach the regime of very wide separations.

The  method  of \citet{TK2016}  is  based  on  the statistics  of  two
quantities:  (i) the angle  $\gamma$ between  the line  joining binary
components (radius vector)  and  the direction  of their instantaneous
orbital motion (vector of relative velocity),  and  (ii) the
normalized  orbital  speed $\mu'$.   Both quantities refer to the
plane of the sky (i.e. are projected) and can be measured for a wide
binary from the position and motion of the secondary component
relative to the primary.  

The characteristic  orbital  speed
$\mu^*$ (in angular units per year) is computed by the formula
\begin{equation}
\mu^* =  (2 \pi \rho)/P^* = 2 \pi \rho^{-1/2} \varpi^{3/2} M^{1/2}, 
\label{eq:mu*}
\end{equation}
where $\rho$ is  the separation between the components, $P^*$ is the notional  orbital period in years
estimated for  a face-on circular orbit from  the projected separation
$s = \rho/\varpi$,  $\varpi$ is the parallax, and $M$  is the mass sum
in solar units. If $\mu$ is  the measured speed of the orbital motion,
$\mu' = \mu/\mu^*$  is its normalized equivalent. A  bound system must
have  $\mu' <  \sqrt{2}$. The  angle $\gamma$  is folded  in  the $(0,
90^\circ)$ interval.

Simulations from \citet{TK2016}  show  that   for  a  thermal  eccentricity  distribution, $\gamma$  is distributed  uniformly and uncorrelated  with  $\mu'$, while
median $\mu' = 0.546$. If $f(e)$ is sub-thermal (i.e. the orbits are,
on average, more circular), the median $\gamma$ increases and becomes
positively correlated with $\mu'$. The fingerprint of a super-thermal
distribution is just the opposite. 

Relevant characteristics of nearby wide binaries and the role of inner
subsystems are covered  in Section~\ref{sec:sample}.  The eccentricity
distribution is addressed  in Section~\ref{sec:e}, first qualitatively
by  examining  the  distributions  of  $\mu'$ and  $\gamma$,  then  by
inverting these distributions to derive $f(e)$. Statistical constraints
on the eccentricity  of any given wide binary that can be deduced from its motion
are outlined.  Section~\ref{sec:disc} discusses  the results
in the context of binary formation.

%%--------------------------------------------------------------
\section{Characteristics of the sample}
\label{sec:sample}

\subsection{Catalogue properties}

The catalogue  of ER2018 was  recovered from the  journal web site  as a
comma-separated  text file.   Only 3601  pairs of  main-sequence stars
with both parallaxes  exceeding 15\,mas are kept in  our subset of the
catalogue.   The  rationale  for  selecting  only nearby  pairs  is  the
increased  accuracy  of the  orbital  speed measurement  (particularly
relevant for  the widest pairs), better screening  for subsystems, and
access  to low-mass,  closer pairs.  At larger  distances,  the ER2018
catalogue suffers from the increasing incompleteness.

\begin{figure}
\centerline{
\includegraphics[width=8.5cm]{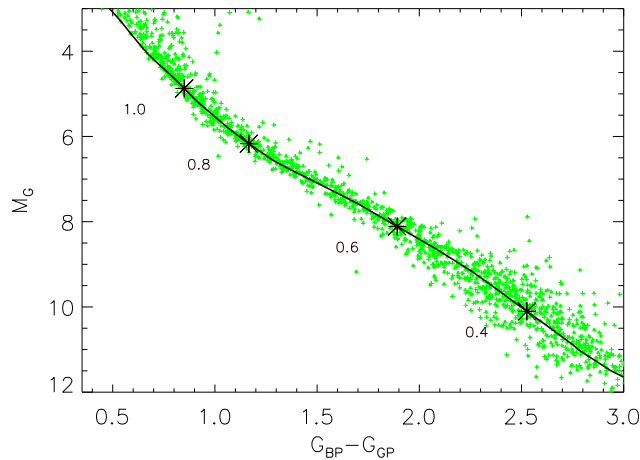} % cmd1.ps
}
\caption{Color-magnitude diagram for  the primary (brighter) components of
  wide   binaries  within  67\,pc.    The  dark   line  is   a  1-Gyr
  solar-metallicity isochrone  \citep{PARSEC}, where the  asterisks and
  numbers mark masses.
\label{fig:cmd} 
}
\end{figure}

Figure~\ref{fig:cmd}  shows the color-magnitude  diagram (CMD) of the
primary components in the
{\it Gaia}  colors.  The 1-Gyr PARSEC  isochrone \citep{PARSEC} traces
the  main  sequence. Masses  are  estimated  from  this isochrone  and
absolute $G$  magnitudes.  Some stars  with masses above 1  \msun ~are
evolved.  To avoid evolved  systems, the statistical analysis below is
restricted to primary  masses less than 1 \msun,  rejecting about 20\%
of the more  massive pairs.  Most pairs have  primary components of G,
K, M spectral types, with a  lowest mass of about 0.2 \msun ~and a
median  mass  of  0.6  \msun.   Figure~18 of  \citet{EB2018}  shows  a
binary-star sequence  in the  CMD which is  practically absent  in the
sample studied here.   The likely reason is a a better  rejection of inner
subsystems in the nearby wide pairs.

\begin{figure}
\centerline{
\includegraphics[width=8.5cm]{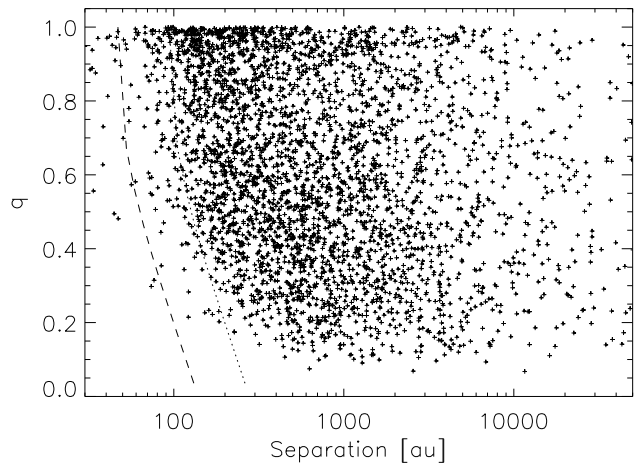} } % sepq.ps
\caption{Correlation  between separation and  mass ratio.   The dashed
  line is the {\it Gaia} detection limit at 67\,pc, the dotted line is
  twice the limit.
\label{fig:sepq} 
}
\end{figure}

Most pairs  have separations  between 100 and  a few thousand  au; the
median separation is 460 au, corresponding to an orbital period of
$\sim$10 kyr.  The ER2018 catalogue has a minimum angular
separation of $2''$  that translates to 133 au  at 67\,pc.  Therefore,
pairs  with $s  <  133$ au come from a smaller volume and their
number is reduced  substantially,
preventing  analysis of closer  binaries.  The  maximum of  the binary
separation  distribution at  $\sim$50 au  or less  \citep{R10}  is not
sampled by this catalogue. Consequently, I focus only on wide pairs.

Correlation between the mass ratio $q = M_2/M_1$ and the separation is
illustrated in  Fig.~\ref{fig:sepq}: pairs with larger $q$  tend to be
closer and, conversely, wide pairs seem to prefer smaller $q$. Please,
refer to  \citet{EB2019} for a  more detailed study of  the mass-ratio
dependence on  primary mass  and separation.  The  dashed line  is the
approximate binary  detection limit  in {\it Gaia},  $\Delta G  < 5.5[
  (\rho/1") - 0.7]^{0.4}$ at the maximum distance of 67\,pc, converted
into $q$ by the relation $q \approx  1 - 0.16 \Delta G$ valid for $M_1
\sim 0.6$  \msun \citep[see Fig.~8 of][]{EB2019}.  The  dotted line is
twice  this  separation.    Figure~\ref{fig:sepq}  suggests  that  the
criteria imposed  in the creation of the  ER2018 catalogue effectively
remove pairs  with separations less  than $\sim$2 times  the detection
limit.

\subsection{Multiple systems}

The  ER2018  catalogue  is  not  fully representative  of  the  unbiased
population of wide  binaries for two main reasons.   First, members of
moving  groups  and clusters  are  excluded.   Second, the  selection
method (good-quality astrometry and photometry, matching proper
motions)  creates  a strong  bias  against  subsystems. Remember  that
sub-arcsecond pairs  often lack {\it  Gaia} astrometry, hence  have no
chance  to be present  in the  catalogue as  components of  wider pairs.
When {\it Gaia} does provide astrometry  of both stars, it can still be
distorted by subsystems (large errors and/or biased PMs), eliminating
wide pairs containing subsystems from the catalogue. However, many
subsystems still remain.  

The ER2018 catalogue  was cross-identified with the 2018  version of the
Washington  Double Star (WDS)  catalogue \citep{WDS}  to look  for known
subsystems  around primary  and secondary  components.   The resulting
list  of 419  candidates was  examined  manually and  compared to  the
Multiple Star  Catalogue, MSC \citep{MSC},  version of July  2019.  Most
hierarchies were  already present  in the MSC,  the missing  ones were
added.  It is  well known that WDS contains  optical and spurious
pairs,  and the  cross-list counts  many such  cases. Decision  on the
reality of each  subsystem, if not obvious, is  based on the available
data and my experience, and  in some cases can be questioned. Overall,
I found that 166 pairs in the ER2018 catalogue within 67\,pc contain known
subsystems.

To  further explore  hidden  multiplicity,  I used  the  list of  {\it
  Hipparcos}     stars    with     astrometric     accelerations    by
\citet{Brandt2018}  and  matched stars  within  67\,pc  with the  WDS.
Systems where the estimated period of the wide pair exceeds $\sim$1000
yr are triple.  Several hundred new triples were added to the MSC as a
result of this effort.  After  this update, comparison of the MSC with the
ER2018 catalogue within 67\,pc reveals 226  multiples, i.e.  60 additional hierarchies
detected by  acceleration.  The total  fraction of known  multiples in
the catalogue  is 0.063. In the following  analysis, hierarchical systems
(i.e.  wide pairs with inner subsystems) present in the MSC are marked
by  a   flag.   Obviously,  there  remain   many  unknown  subsystems,
especially  among fainter  stars not  covered by  {\it  Hipparcos} and
\citet{Brandt2018}.   \citet{EB2018}  estimate that  as  much as  0.36
fraction  of their sample  could contain  a subsystem  in one  or both
components,  despite  the  bias  against multiples  inherent  to  this
catalogue. The fraction of undetected subsystems in the subset of nearby wide
binaries studied here should be less than in the full catalogue.

%---------------------------------------------------------
\section{Eccentricity distribution}
\label{sec:e}

\subsection{Qualitative view}

Parameters of the relative motion  of wide pairs are computed from the
{\it Gaia}  astrometry provided in the ER2018  catalogue.  The modulus
of the relative  proper motion (PM) $\mu$ is computed  from the PMs of
each component,  and its  error is computed  by eq.~6 in  ER2018.  The
separation $\rho$ and position angle $\theta$ of the pair are computed
from  the  components' coordinates.   The  angle  of  the relative  PM
$\theta_\mu$ is computed similarly. The angle $\gamma$ is evaluated in
three steps: $\gamma_1 = | \theta - \theta_\mu|$, $\gamma_2 = \gamma_1
\; {\rm  modulo} \;  180^\circ$, $\gamma =  \gamma_2$ for  $\gamma_2 <
90^\circ$ and  $\gamma = 90^\circ  -\gamma_2$ otherwise. The  error of
$\gamma$  is determined  by the  relative PM  error, and  expressed in
degrees:  $\sigma_\gamma =  57.3  \; \sigma_\mu/  \mu$,  based on  the
assumption  that  PM  errors  are isotropic.   The  characteristic  PM
$\mu^*$ is  computed using (\ref{eq:mu*}).  The  calculation of $\mu'$
accounts  for  a  small  bias  due  to  measurement  errors:  $\mu'  =
\sqrt{\mu^2 - \sigma_\mu^2}/\mu^*$.   However, the results are similar
if this correction is neglected.

The following statistics use wide pairs satisfying the following conditions:
\begin{itemize}
\item
$M_1 < 1$ \msun 

\item
$\sigma_\gamma < 50^\circ$

\item
$\mu' < 1.41$ 

\item
Not multiple
\end{itemize}
Filtering leaves 2663 pairs for further analysis.  The large tolerance
on  $\sigma_\gamma$ is  intentional  to avoid  potential  bias in  the
statistics  caused  by rejection  of  slowly  moving  pairs.  Only  99
(3.7\%) of  the 2663 accepted  pairs have $20^\circ <  \sigma_\gamma <
50^\circ$.   Very  similar  results  are obtained  with  the  stricter
$\sigma_\gamma < 20^\circ$ cut.

\begin{figure}
\centerline{
\includegraphics[width=8.5cm]{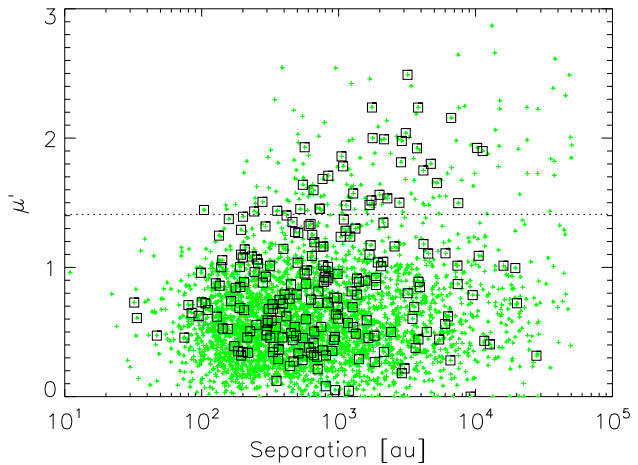} % mu-sep.ps
}
\caption{Statistics  of  ``fast  movers'':  dependence  of  $\mu'$  on
  projected separation.   Known multiples  are marked by  squares, the
  dotted line marks $\mu' = 1.41$.
\label{fig:fast} 
}
\end{figure}
\begin{figure}
\centerline{
\includegraphics[width=8.5cm]{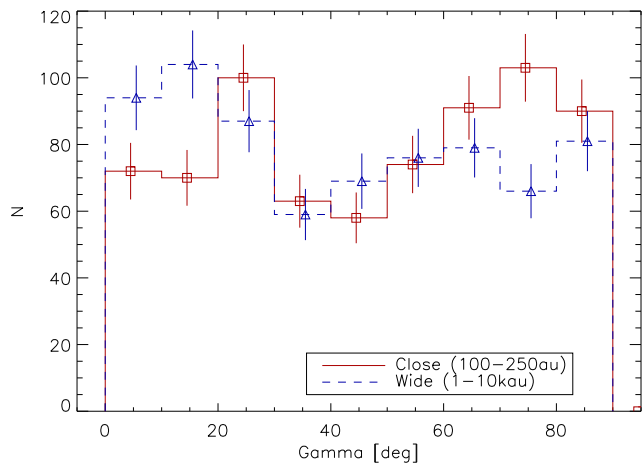} % gamhist.ps
}
\caption{Histograms of $\gamma$ for close
  ($100  < s  <  250$ au, $N=721$)  and wide  ($10^3  < s  < 10^4$  au,
  $N=715$)  pairs. 
\label{fig:gamma} 
}
\end{figure}

Apparently  unbound pairs  with  $\mu' >  1.41$,  or ``fast  movers'',
deserve a closer look.  Their total number is 179, or 0.05 fraction of
all pairs.  Reasons such  as measurement errors or hidden multiplicity
reveal  themselves  in  the  statistics  of fast  movers  explored  in
Fig.~\ref{fig:fast}.  The widest (hence slower moving) pairs should be
more affected by  these factors.  Indeed, the fraction  of fast movers
increases with separation: it  is 86/1086=0.079 for $s$ between $10^3$
and $10^4$  au and 29/157=0.185 for  $s>10^4$ au.  On  the other hand,
the fraction  of fast  movers does not  depend on the  distance.  This
suggests that measurement errors are  not the dominant factor and that
fast movers are mostly caused by motions of inner subsystems.  Indeed,
the fraction of known multiples among the fast movers, 37/179=0.21, is
much  larger than  the fraction  of multiples  in the  full catalogue,
0.06.  \citet[][Sect. 3.5]{Belokurov2020}  prove that excessive motion
of some  wide pairs in the  ER2018 catalogue is  related to unresolved
inner subsystems, rather than to  a modified gravity law, as suggested
by \citet{Pittordis2019}.

\begin{table}
\centering
\caption{Medians of $\mu'$ and $\gamma$ and correlation $C_{\mu' \gamma}$ }
\label{tab:med}
\medskip
\begin{tabular}{l c c c c  }
\hline
Sample  & $N$  & $\gamma_{\rm med}$ & $\mu'_{\rm med}$  & $C_{\mu' \gamma}$ \\
\hline
(1) All           & 2663  & 44.7$\pm$1.0 & 0.542 & $-$0.02 \\
(2) 100--200 au    &  515  & 51.2$\pm$2.1 & 0.531 & 0.03 \\
(3) 200--400 au &    598 & 43.7$\pm$2.9  & 0.523 & $-$0.09 \\  
(4) $10^3-10^4$ au & 715  & 41.5$\pm$2.2 & 0.554 &  0.01 \\
(5) Multiple      &  226 & 43.9$\pm$2.6 & 0.842 & $-$0.06 \\
\hline 
$f(e)=2e$     &  --- &  45.0 &  0.546 & 0.00 \\
$f(e)=1 $     & ---  &  54.1 &  0.606 & 0.18 \\
$e=0$         & ---  &  68.3 &  0.677 & 0.61 \\
\hline
\end{tabular}
\end{table}

Table~\ref{tab:med} gives the median values of $\gamma$ and $\mu'$ and
their correlation  coefficient $C_{\mu'  \gamma}$ for the  full sample
(1),  subsamples  (2)-(4)  selected  according projected  physical  to
separation, and the control sample  (5) of pairs with subsystems.  The
errors of the  $\gamma$ medians are estimated by  bootstrap.  The last
lines give  the theoretical  parameters computed by  \citet{TK2016} by
simulation  of thermal  and  flat eccentricity  distributions and  for
circular orbits.  The statistics of multiples (5), excluded from other
samples, illustrate the influence of subsystems.  They increase $\mu'$
by  adding kinematic  ``noise'' and  bias  the mass  sum that  affects
calculation  of $\mu^*$. Indeed,  the median  $\mu'$ for  multiples is
larger than in other samples.  Motions caused by the subsystems should
also  randomize  $\gamma$.  Unrecognized  multiples  remaining in  the
sample might bias the results to some extent, especially at large $s$.

\begin{figure*}
\centerline{
\includegraphics[width=16cm]{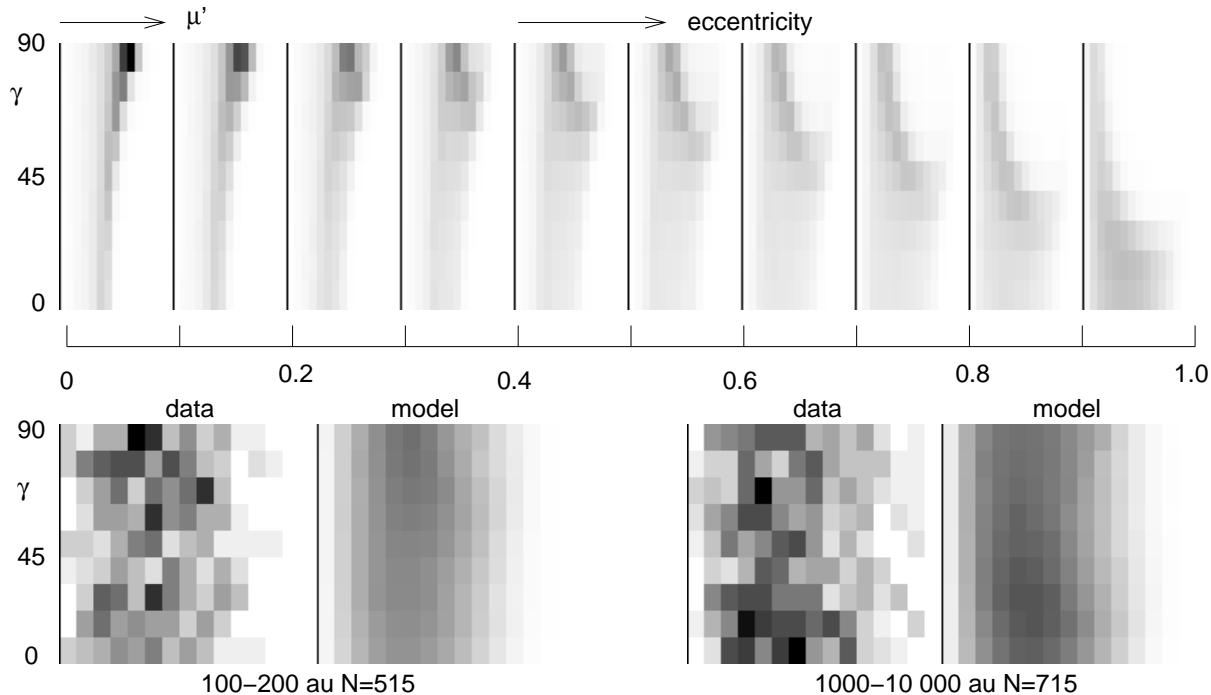} } % modeling.eps
\caption{Restoration of the eccentricity distribution.
   Top row: ten templates in order of increasing
  eccentricity. Bottom row: observed histograms and their models. 
\label{fig:modeling} 
}
\end{figure*}

Table~\ref{tab:med}  demonstrates  that  the  statistics  of  relative
motion  in the  full  sample  (1) are  perfectly  compatible with  the
thermal eccentricity  distribution.  However, the  sub-samples (2) and
(4)  (close and  wide pairs,  respectively) deviate  from  the uniform
distribution of $\gamma$ in opposite ways.  The difference between the
$\gamma$ medians between these sub-samples, $9.6^\circ \pm 2.8^\circ$,
is   statistically  significant.   Figure~\ref{fig:gamma}   shows  the
distributions of $\gamma$  for close (100-250 au) and  wide (1-10 kau)
pairs,  with a similar  number of  pairs in  each group.   Close pairs
appear  to  be sub-thermal  (less  eccentric),  wide  pairs suggest  a
super-thermal eccentricity  distribution.  Wide pairs seem  to have an
excess of  $\gamma <  30^\circ$ values.  Interestingly,  the histogram
for  the close  binaries might  contain  a similar  excess, while  the
remaining  pairs  have  a  $\gamma$  distribution  increasing  towards
$\gamma  =   90^\circ$,  indicative  of   a  sub-thermal  eccentricity
distribution.

The parameter $\mu'$ depends on the estimated masses and  is
biased to larger values by undetected subsystems. On the other hand,
the angle $\gamma$ is a purely geometric parameter. Both the measurement
errors and the additional motions caused by subsystems can only smooth
the true  distribution of $\gamma$. Therefore, the deviation of
$\gamma$ from the uniform distribution evidenced by
Table~\ref{tab:med}  is a robust, albeit qualitative, indicator of the
non-thermal eccentricity distribution.

When the  sample is split into  groups according to the  mass ratio or
primary  mass,  the  resulting   distributions  of  $\gamma$  show  no
difference. However,  when I compare  122 wide twins  ($q>0.95$) with
$s>500$ au  with 213 similarly  wide pairs with  $0.8 < q  <0.95$, the
median  $\gamma$  values are  $36.7\pm3.6$  and $44.0\pm3.9$  degrees,
respectively,  suggesting that  wide twins  might have  more eccentric
orbits.

%---------------------------------------------------------
\subsection{Recovering the eccentricity distribution}
\label{sec:fe}

In this Section the  eccentricity distribution $f(e)$ is inferred from
the  joint  distribution  of   $\mu',  \gamma$  using  the  method  of
\citet{TK2016},  briefly  recalled  here.   The histogram  of  $(\mu',
\gamma)$ is computed on a 15$\times$9 grid (0.1 bin size in $\mu'$ and
$10^\circ$  bins  in  $\gamma$).   Such  distributions  for  simulated
binaries with  eccentricity in  a narrow range,  e.g.  from 0  to 0.1,
constitute  a  set  of  theoretical distributions  (templates).   Each
template is  evaluated for  $10^4$ simulated binaries.   Parameters of
simulated  binaries such  as mass  sum, parallax,  and  separation are
sampled randomly from the real  data, and appropriate cuts are applied
($\rho >2''$, $s >100$  au, $\sigma_\gamma < 50^\circ$). This accounts
for potential biases in the real sample.

The observed distribution  of $(\mu', \gamma)$ is modeled  by a linear
combination  of templates  with 10  coefficients $f_k$  satisfying the
constraint $\sum_k f_k  = 1$. Let $n_{i,j}$ be  the number of binaries
in  the $i$-th and  $j$-th bin  of the  histogram ($i$  corresponds to
$\mu'$  and $j$  to $\gamma$),  $T_{i,j,k}$  -- the  template for  the
$k$-th eccentricity  bin, normalized to a  unit sum over  $i$ and $j$.
The model is $n'_{i,j} = N  \sum_k T_{i,j,k} f_k$, where $N$ is number
of  binaries in the sample. The  goodness  of  fit is  quantified  by the  parameter
$\chi^2 = \sum_{i,j}  (n_{i,j} - n'_{i,j})^2/\sigma^2_{i,j}$, assuming
Poisson errors $\sigma^2  _{i,j} = n_{i,j}$ and $\sigma^2  _{i,j} =1 $
if $ n_{i,j} =0$.

%\input{fetable.tex}
% Eccentricity dstributions
\begin{table*}
\centering
\caption{Eccentricity distribution}
\label{tab:fe}
\medskip
\begin{tabular}{l ccc ccc ccc ccc c  }
\hline
Separation  & $N$  & $\langle e \rangle$ & $\chi^2/126$ & 
$f_1$ &$f_2$ &  $f_3$ &$f_4$ &$f_5$ &$f_6$ &$f_7$ &$f_8$ &$f_9$ &$f_{10}$ \\
\hline
100--10$^4$ au & 2463 & 0.684 &  1.22 &   0.002& 0.033& 0.046& 0.053& 0.102& 0.128& 0.095& 0.134& 0.169& 0.238 \\ % 0.001
100--200 au &   515   & 0.639 & 1.31  &   0.018& 0.024& 0.040& 0.069& 0.113& 0.148& 0.155& 0.145& 0.139& 0.148 \\ % 0.01
100--200 au$^{\rm a}$ &   515   & {\it 0.643} & {\it 1.22}  & {\it   0.021}&{\it  0.026}&{\it  0.038}& {\it 0.067}&{\it  0.111}&{\it  0.144}&{\it 0.149}& {\it0.142}&{\it 0.144}&{\it 0.159} \\ % 0.01 alt. grid
200--10$^3$ au & 1233 & 0.677 & 1.25  &   0.000& 0.033& 0.049& 0.059& 0.092& 0.124& 0.119& 0.140& 0.171& 0.213 \\ % 0.002
10$^3$--10$^4$ au& 715& 0.683 & 1.37  &  0.023& 0.029& 0.040& 0.059& 0.080& 0.096& 0.113& 0.147& 0.188& 0.227 \\ % 0.02
\hline
\end{tabular}
$^{\rm a}$~Alternative templates generated using only 
  binaries with $s = 100-200$ au  are used.
\end{table*}

A  small smoothing  parameter $\alpha$  is  introduced  to favor  continuous
distributions and  effectively damp the  noise in the  resulting $f_k$. 
The distribution $f_k$ is found by minimizing 
\begin{equation}
\sum_{i,j} (n_{i,j} - n'_{i,j})^2 + \alpha \sum_k (f_k - f_{k+1})^2 \rightarrow {\rm min}. 
\label{eq:min}
\end{equation}
with the constraint  $\sum_k f_k  = 1$. A  linear equation can  be derived
from  (\ref{eq:min}),   yielding  the  solution  $f_k$   as  a  simple
matrix-vector  product.   Note  that  the minimization  does  not  use
weights  corresponding  to  $\sigma_{i,j}$  for reasons  explained  by
\citet{TK2016}, namely to reduce the influence of small $n_{i,j}$ that
can  be biased,  e.g.  by  fast movers,  and to give  more weight  to the
histogram bins  with large values.  Therefore, the solution of
  (\ref{eq:min}) does not correspond to the  $\chi^2$ minimum. 
   Compared to \citet{TK2016},   the regularization  term is
changed from $\alpha \sum_k f_k^2$  to the sum of squared differences.
The new  formulation models the  data by a smooth  distribution $f_k$, while
the  previous   formulation  biased  the  result   towards  a  uniform
distribution.    However,  the  results  delivered by  those   two  alternative
regularization schemes are almost identical.

Adequacy of  the model  is confirmed by  checking that  the normalized
values of  $\chi^2/(135 - 9)$ are close  to one (135 is  the number of
bins,  9 is  the number  of  degrees of  freedom).  For  9 degrees  of
freedom, the  ``1 $\sigma$'' confidence limit  (68.3\%) corresponds to
the hyper-volume  in the parameter  space where $\chi^2$  increases by
less   than  10.4   relative  to   its  minimum   \citep{Press}.   The
regularization parameter  $\alpha$ is increased from  $10^{-3}$ with a
step of  2 times until the  $\chi^2$ exceeds its minimum  value by more
than  10.4; then  the previous  (one step  back) $\alpha$  is adopted.
Typically, $\alpha \approx 10^{-2}$ for $N \sim 500$.

The  solution of (\ref{eq:min}) does  not guarantee  that all $f_k  \ge 0$.
When a negative  $f_k$ is encountered, it  is set to zero and  the fit is
repeated with the reduced number of free parameters (so-called non-negative
least squares).

Figure~\ref{fig:modeling}  illustrates  the  method. The  upper  panel
shows  10 template  distributions, all  on  the same  gray scale.  The
vertical axis is $\gamma$, the  horizontal axis is $\mu'$ ranging from 0 to
1.5  in each  template. At  small  eccentricity, the  maximum is  near
$\gamma  = 90^\circ$,  $\mu'=1$, while  small $\gamma$  dominate for
eccentric orbits.  The  lower panels show the observed distributions
in two  separation ranges and their models.

\begin{figure}
\centerline{
\includegraphics[width=8.5cm]{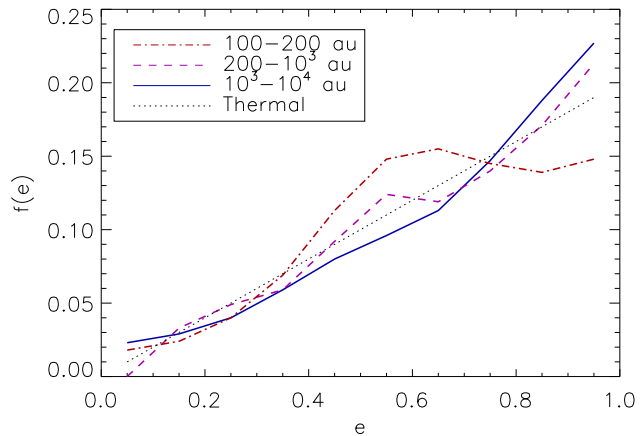} } % feplot.ps
\caption{Eccentricity distributions. 
\label{fig:fe} 
}
\end{figure}

The  method was  applied to  the full  sample and  to various  cuts in
separation. Representative results are given in Table~\ref{tab:fe} and
plotted in Fig.~\ref{fig:fe}. Columns 3 and 4  of Table~\ref{tab:fe} give
the  mean eccentricity  $\langle e  \rangle$ and  the goodness  of fit
metric $\chi^2/126$, which  is larger than one because  no weights are
used.  At 200-10$^3$ au separations,  $f(e)$ is practically thermal. At
separations below  200 au the deficit of  large eccentricities becomes
apparent,  while at   separations of 1-10 kau $f(e)$  becomes slightly
super-thermal. However, undetected subsystems mostly affect $f(e)$ at the largest
separations, so the result should be taken with some caution.

At the shortest  separations, the cuts applied to  the data affect the
statistics because the closest  and fastest-moving binaries might fall
below the  $2''$ cutoff in  $\rho$.  The templates were  re-computed by
selecting  only pairs  with $s  < 200$  au from  the real  sample and,
indeed, they differ from the templates obtained using the full sample,
athough they  look qualitatively similar.  For the  100-200 au sample,
the calculation  was repeated  using those more  appropriate templates
(see the  numbers in italics in Table~\ref{tab:fe}).   Indeed, a lower
$\chi^2$  was  obtained,  but  the  resulting $f_k$  differ  from  the
standard calculation by no more than 0.01.

\subsection{Posterior eccentricity distribution}
\label{sec:posterior}

In some  cases, constraints on  the eccentricity of a  particular wide
binary  are useful,  e.g. to  find the minimum  periastron  separation and
evaluate the influence of the wide companion on an inner subsystem or
a circumstellar disk. In response to this need, sampling techniques like
``Orbits for the Impatient'' (OFTI) were developed to derive posterior
distributions   of    orbital   parameters   from    incomplete   data
\citep{OFTI}. A simpler  and faster way to relate  the two observables
$x_0  = (\mu',  \gamma)$ with the posterior eccentricity  distribution is
outlined here. The joint distribution of $x$ and $e$ can be written as
\begin{equation}
f(e, x) = f_1(x | e) \; f(e) .
\label{eq:fex}
\end{equation}
If $f(e)$ in the right-hand part is replaced by the prior distribution
$f_0(e)$  and  $x$ takes  the  observed  value  $x_0$, we  obtain  the
posterior  distribution  $f(e |  x_0)$.  The conditional  distribution
$f_1(x  |e)$ is  obtained from  simulations and  is equivalent  to the
templates $T_{i,j,k}$ with suitable re-normalization.

\begin{figure}
\centerline{
\includegraphics[width=8.5cm]{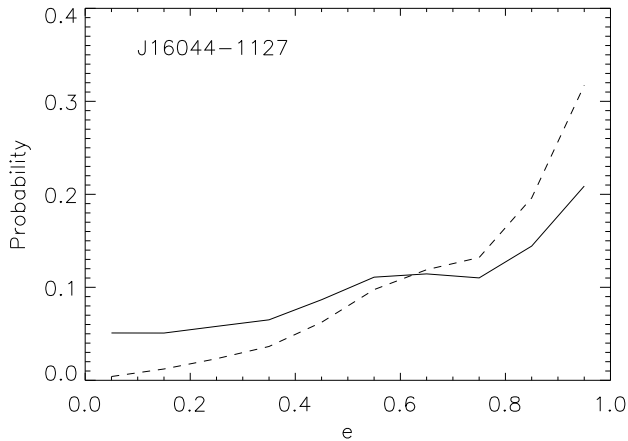} % eposterior.ps
}
\caption{Posterior eccentricity distributions of ADS~9910 (WDS
  J16044$-$1127).
Full line -- uniform prior, dashed line -- linear prior.
\label{fig:ADS9910} }
\end{figure}

To   give   an   example,   Fig.~\ref{fig:ADS9910}   plots   the posterior
eccentricity distributions  of a  typical nearby wide  binary ADS~9910
($s  = 335$  au, $P^*  = 4.5$  kyr) corresponding  to the  uniform and
linear prior distributions $f_0(e)$. The observed parameters are $\mu'
= 0.33$ and $\gamma =  30^\circ$, and there are no inner subsystems. Although
a large eccentricity is  more likely, any eccentricity  is possible, hence
this posterior constraint  is fuzzy.  However,  in other instances  the posterior
constraints  can be  stronger and  less  dependent on  the prior.  For
example, $\mu'  \le 1$ for circular orbits, so an observed value $\mu'
> 1$   immediately   excludes   circular  orbits.   Similarly,   large
eccentricities can be ruled out for certain combinations of $\mu'$ and
$\gamma$.

The posterior  distributions of  orbital parameters
such as eccentricity depend  on the adopted prior distributions, i.e.
are model-dependent.  In the sampling  methods like OFTI,  the derived
posterior distributions depend  on the distribution of samples in
the multi-dimensional  space of  orbital parameters. This  work proves
that a  linear eccentricity distribution is the  recommended prior for
wide  binaries,  while a  uniform  prior  should be deprecated in  this
context.

%---------------------------------------------------------
\section{Discussion}
\label{sec:disc}

In \S~5.1.4  of their review on  multiplicity statistics, \citet{DK13}
wrote that observations provide a ``clear and uniform picture'' of the
eccentricity  distribution, but  in  the same  paragraph propose  two
different models of $f(e)$, flat and Gaussian. Their discussion refers
mostly  to  spectroscopic binaries.   Figure~15  of \citet{R10}  gives
$f(e)$ for visual binaries that  is approximately flat between 0.1 and
0.6  and drops  at large  and small  $e$. Based  on this  result, some
authors  adopted a  flat  $f(e)$ between  0  and 0.8  and $f(e)=0$  at
$e>0.8$.

\citet{TK2016}  derived  $f(e)$ than  rises  linearly  at $e<0.8$  and
remains constant at larger $e$.  The median separation in their sample
was 120 au.   This result is confirmed here  using a different, larger
sample  (red  curve  in  Fig.~\ref{fig:fe}).  They also  demonstrated  a  good
agreement between the reconstructed eccentricity distribution and
  $f(e)$ derived from known visual orbits with $P > 100$ yr,
except for the  last bin  lacking eccentric orbits.  The
paucity of binaries with $e>0.8$  in the sample of \citet{R10} is explained
by the bias in the orbit catalogue, while in fact such pairs are frequent.

At  separations  $s>200$  au,  the eccentricity  distribution  becomes
consistent with the thermal one. Moreover, this study shows that pairs
with  $s> 1$  kau  might have  a  slightly super-thermal  eccentricity
distribution.   The  $f(e)$  derived  here for  these  binaries  could
potentially  be  biased   by  undiscovered  subsystems.  However,  the
statistics of  $\gamma$, less affected  by subsystems, also  suggest a
super-thermal eccentricity distribution at  $s>1$ kau.  It is possible
that the  trend to super-thermal  $f(e)$ at large separations  is even
stronger than found here.

A close match of the observed eccentricity distribution to the thermal
distribution  suggests that  dynamical processes  played  an important
role in the formation of wide binaries. For example, pairs that remain
after  dynamical interaction  of  a  binary or  a  triple system  with
another   star    (scattering)   have   a    nearly   thermal   $f(e)$
\citep[e.g. Fig. 16  in][]{Antognini2016}.  When two binaries interact
dynamically  in  a gas  cloud  and survive  as  a  2+2 quadruple,  the
eccentricity  of   the  wider  remaining  pair   has  a  super-thermal
distribution, while  for the closer  pair the distribution  is thermal
\citep[Fig.  9  of][]{Ryu2017}.   Disintegration  of  unstable  triple
systems   leaves   binaries  with   a   mildly  super-thermal   $f(e)$
\citep{Stone2019}.   However, wide  binaries formed  by  ejection from
unstable triples or  ``unfolding'' \citep{Reipurth2012} must have only
very eccentric orbits.  The unfolding mechanism does not match certain
properties of real wide binaries  and appears to be exceptional rather
than typical \citep{Tok2017}.

Hydrodynamical  simulations  show  that  protostars  formed  at  large
distances from each other can get bound into a wide pair with the help
of  gas friction \citep{Bate2019,Kuffmeier2019}.   Continued accretion
onto  such binary shortens  its period  and reduces  the eccentricity.
From this  perspective, eccentric  orbits are produced  naturally even
without   dynamical  interactions   with  other   stars;   a  positive
correlation  between separation  and eccentricity  is  expected.  This
formation  channel   of  wide  binaries  works   even  in  low-density
environments.  However,  no predictions of  the resulting eccentricity
distribution are available so far.   Simulations of a dense cluster by
\citet{Bate2019}  inform  us on  the   binary  statistics,  but  in  this
environment   dynamical  interactions between stars are  important  and  the
resultant statistics reflect a complex interplay of several processes.

To disentangle the relative  roles of stellar dynamics and gas-assisted
capture in the  formation of wide binaries, it  will be interesting to
apply  this method to  sparse associations  like Taurus-Auriga  and to
moving  groups.  Small sample  size  and the need to  account  for the  subsystems
present obvious challenges to this endeavor.

\section*{Acknowledgments}
I thank Karim El-Badry and Maxwell Moe for comments on the early
version of this work  and the anonymous Referee for helping to
  improve the presentation. 
This study used the Washington  Double Star Catalogue maintained at USNO.
It has made  use of data from the European  Space Agency (ESA) mission
{\it  Gaia} ({\it https://www.cosmos.esa.int/gaia}), processed  by the
{\it   Gaia}   Data   Processing   and  Analysis   Consortium   (DPAC,
{\it https://www.cosmos.esa.int/web/gaia/dpac/consortium}).     Funding
for the DPAC has been provided by national institutions, in particular
the  institutions   participating  in  the   {\it  Gaia}  Multilateral
Agreement.

%%--------------------------------------------------------------
%\section{}
%\label{sec:}

\section{Supplementary material}
The  supplementary file  lists  $\mu$, $\gamma$, $\mu^*$, and other  parameters
described in the text for 3601 wide pairs from the ER2018 catalogue located
within 67\,pc. 

\label{lastpage}

\end{document}